\documentclass[aps,pra,reprint,twocolumn,superscriptaddress,floatfix
]{revtex4-1}
\usepackage{graphicx}
\usepackage{dcolumn}
\usepackage{amsmath}
\usepackage{amsfonts}
\usepackage{epstopdf}
\usepackage{euscript}
\usepackage{subfigure}
\usepackage[usenames]{xcolor}

\begin{document}
\title{Selective Bond-Breaking in Formic Acid by Dissociative Electron Attachment}
\author{D. S. Slaughter}
\affiliation{Lawrence Berkeley National Laboratory, Chemical Sciences, Berkeley, California 94720, USA}
\author{Th. Weber}
\affiliation{Lawrence Berkeley National Laboratory, Chemical Sciences, Berkeley, California 94720, USA}
\author{A. Belkacem}
\affiliation{Lawrence Berkeley National Laboratory, Chemical Sciences, Berkeley, California 94720, USA}
\author{C. S. Trevisan}
\affiliation{Department of Sciences and Mathematics, California Maritime Academy, Vallejo, California 94590, USA}
\author{R. R. Lucchese}
\affiliation{Lawrence Berkeley National Laboratory, Chemical Sciences, Berkeley, California 94720, USA}
\author{C. W. McCurdy}
\affiliation{Department of  Chemistry, University of California, Davis,  California 95616, USA}
\affiliation{Lawrence Berkeley National Laboratory, Chemical Sciences, Berkeley, California 94720, USA}
\author{T. N. Rescigno}
\affiliation{Lawrence Berkeley National Laboratory, Chemical Sciences, Berkeley, California 94720, USA}

\date{\today}

\begin{abstract}
We report the results of a joint experimental {and} theoretical study of dissociative electron attachment to formic acid (HCOOH) in the 6-9 eV region, where H$^-$ fragment ions are a dominant product. Breaking of the CH and OH bonds is distinguished experimentally by deuteration of either site. We show that in this region H$^-$ ions can be produced by formation of two {or possibly three} Feshbach resonance (doubly-excited anion) states, one of which leads to either C-H or O-H bond scission, while the other can only produce formyloxyl radicals by O-H bond scission. Comparison of experimental and theoretical angular distributions of the anion fragment allows the elucidation of state specific pathways to dissociation.
\end{abstract}

\maketitle
\section{Introduction}
Low-energy electrons play a key role in the radiation-induced chemistry of biomolecules, in atmospheric physics and chemistry, and in materials processing and imaging involving ionizing radiation. In particular, dissociative electron attachment (DEA) is one of the fundamental interactions that drive free-electron chemistry and continues to attract considerable interest, not only for the need to understand negative ion production and electron-induced molecular breakup but also to understand and model these processes in systems of technological interest~\cite{Fabrikant}. The dynamics associated with DEA can be complex, involving different fragment ion channels and conical intersections between different anion states~\cite{haxtonh20, Moraco2, ammonia}. DEA to formic acid, the simplest of the organic acids, has significance in many different contexts, including chemistry in planetary atmospheres and in space, precursors or intermediates in various synthetic processes, and in the formation of biologically relevant molecules through the production of reactive radicals due to radiative and charged particle interactions. 

The laboratory frame angular distributions of fragment ions can provide insight into the breakup process, because they help to identify the associated resonance state and can be a key ingredient in unraveling the underlying dynamics.  Comparing  theoretical calculations of those angular distributions with experimental observations is key to making those assignments in the DEA processes we study here and provides a powerful tool for elucidating their mechanisms and dynamics.

Much of the previous work on electron interactions with formic acid, both {theoretical}~\cite{CSTformic, TNRformic, Gallup1, Gallup2}  and experimental~\cite{Pelc, Prab05, Allanformic, bhargavaramVelocityImagingFormic2020}, has focused on the mechanism of electron attachment around 1.8 eV which results in production of H + formate anions (HCOO$^-$). Our focus here is on the incident electron energy range between 6 and 9 eV, where H$^-$ is the predominant ion produced through the formation of doubly-excited transient anion states (Feshbach resonances). Prabhudesai {\it et al.}~\cite{Prab05} were the first to observe H$^-$ from dissociative electron attachment (DEA) to formic acid and to provide absolute values of the DEA cross sections.  Although H$^-$ is the dominant anion produced in the 6-9 eV range, it was not reported in the earlier measurements of Pelc {\it et al.}~\cite{Pelc}, presumably because their quadrupole mass spectrometer was not well-suited to isolate and detect H$^-$.

H$^-$ ions from DEA to formic acid can originate from dissociation of either the C-H bond or the O-H bond:

\begin{figure} [h]
	\includegraphics[width=7.5cm]{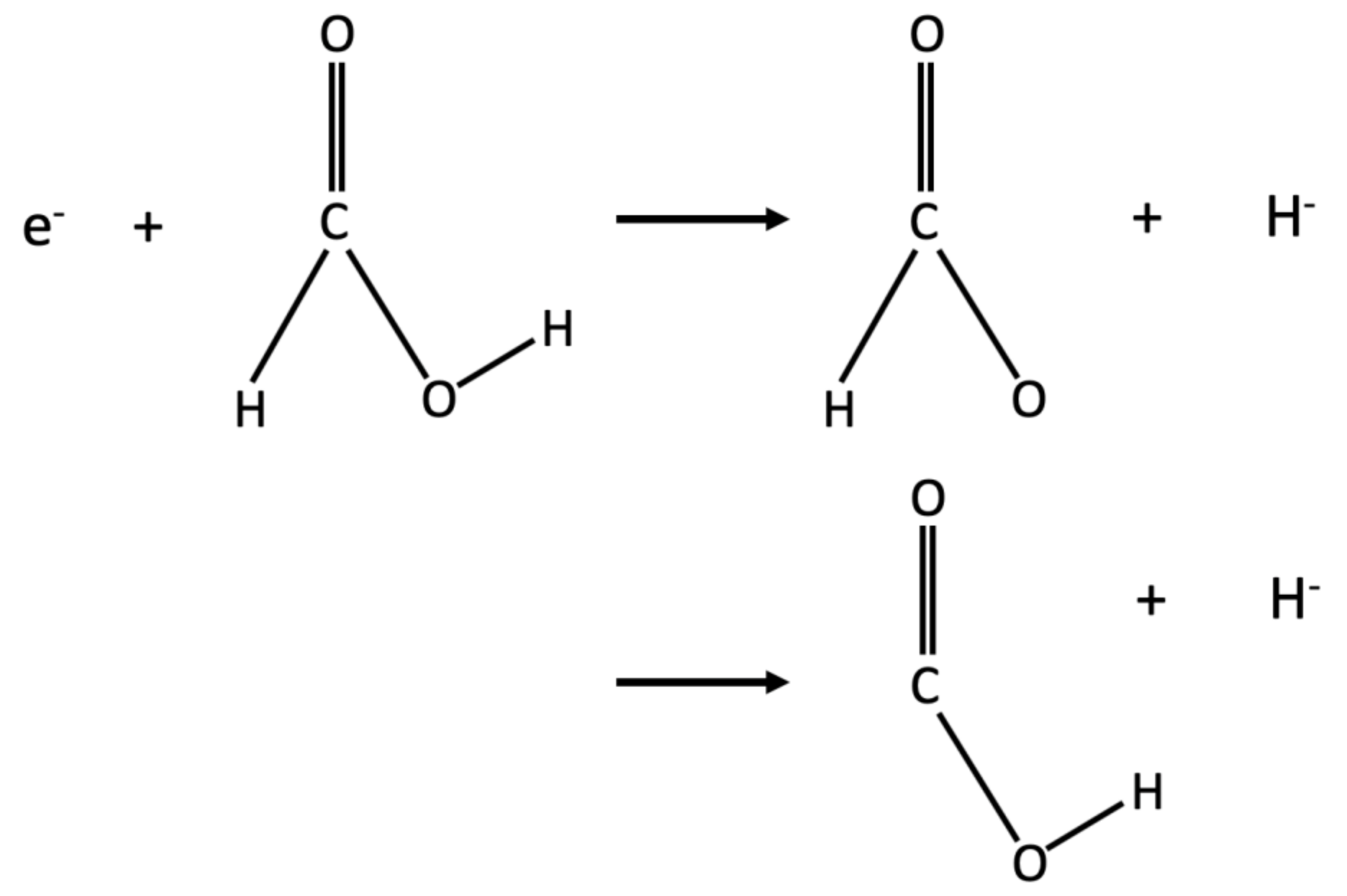}
\end{figure}
A C-H break will result in the formation of the HOCO radical, an important intermediate in atmospheric and combustion chemistry, which has been subject of numerous theoretical and experimental studies~\cite{miller, bowman, continetti, continetti2, miyabe}. An O-H break, on the other hand, produces a formyloxyl radical, HCOO which, in addition to its metastability with respect to H + CO$_2$ dissociation~\cite{neumark}, is characterized by having several low-lying electronic states~\cite{mclean, rauk}. Our objective here is to use a combination of experiment and {\it ab initio} theory to characterize the doubly-excited states that lead to H$^-$ production and elucidate the state-specific pathways to dissociation. By comparing anion fragment yields for two deuterated isotopologues of formic acid, we can distinguish C-H from O-H bond scission. Electronic structure calculations are performed to determine the relevant resonant anion states and their dissociation paths to possible products. We also carry out complex Kohn electron-molecule scattering calculations to determine the expected ion angular distributions, under the assumptions of axial recoil fragmentation, which are compared with measured distributions to confirm the resonant states and product assignments.

We begin with a brief description of the DEA reaction microscope used in the experiment. This is followed by a description of the theoretical methods employed. We proceed to a discussion of the dissociation dynamics and H$^-$ angular distributions and  a comparison of the experimental data with our theoretical predictions. We conclude with a brief discussion.

\section{Experimental Setup}
The yield and momentum of each mass-resolved anion fragment was analyzed using a DEA reaction microscope, which has been described in detail previously~\cite{Adaniya12, topical}, so only the most important details specific to the current work are included here. The experiments employ an energy-tunable, pulsed electron beam that is perpendicular to the time-of-flight axis of a 3-D momentum imaging spectrometer. Deuterated formic acid (HCOOD or DCOOH) is collimated by a long stainless steel capillary of 0.3 mm inner diameter to form the room temperature effusive gas target. The capillary was heated to 120$^\circ$C, to avoid condensation and to maintain the continuous flow of the formic acid vapor. The molecular beam was produced using low pressure conditions, $<$~2~Torr within the tubing upstream of the capillary, to ensure negligible contributions from formic acid dimers~\cite{allan_electron_2007}. The electron gun is pulsed at 50 kHz and, with an adjustable delay following each electron gun pulse, the first electrode of the anion spectrometer is pulsed extracting negative ion fragments into the spectrometer. A uniform magnetic field of typically 25~G, coaxial to the electron beam, allows the separation of anions from the scattered electron background and assists in the low energy electron beam transport and collimation. The electron beam energy spread is 0.5~eV (full width at half maximum) and the electron beam mean energy was calibrated by the O$^-$ onset from CO$_2$ at 3.99~eV with a precision of $\pm$0.1~eV for each experiment.

The 3-D momentum-imaging anion spectrometer consists of weak 12~V/cm anion extraction and acceleration fields and a position-focusing lens, with field transitions established by grid electrodes, to map the momentum of anion fragments onto the position and time-sensitive anion detector. The detector is a pair of 75~mm diamater chevron microchannel plates that amplify each detected particle onto a two-layer delay line anode, allowing for the event-by-event acquisition of the 3-D momentum of each ion, encoded in the time and position of each ion hit. The detector and spectrometer are electrically shielded to prevent most of the background scattered electrons from entering the spectrometer or hitting the detector. In the list-mode data record, the {three-dimensional} momentum of each detected ion fragment is stored, allowing for both on-the-fly and offline analysis.

\section{Theoretical Approach}
The ground state geometry of neutral formic acid is planar and has C$_s$ symmetry. It is nominally described by the electronic configuration (1-8a$'$)$^2$(1a$''$)$^2$(9a$'$)$^2$(2a$''$)$^2$(10a$'$)$^2$. Low-lying excited states, excited by electron collisions, can serve as parents of a doubly-excited state when an electron in an occupied orbital is promoted to an unoccupied orbital and the colliding electron is captured into the same orbital. The process is:
\begin{equation} 
e^-+( \psi _{occ}) ^2=\psi _{occ}(\psi _{unocc})^2
\end{equation}
The first excited electronic state of formic acid is a $^3$A$''$ state, corresponding to the excitation 10a$'$ $\rightarrow$ 3a$''$ (here denoted by n$_0$ $\rightarrow$ $\pi^*$). The 3a$''$ orbital is a compact, anti-bonding  valence orbital, which is responsible for the 1.8 eV shape resonance seen in low-energy elastic scattering~\cite{CSTformic}. Higher energy resonances are expected to involve valence electronic excitation of the target molecule. This is the case for several other systems we have studied (H$_2$O, CO$_2$, CH$_3$OH) that exhibit narrow Feshbach resonances, having double occupation of a $\sigma^*$ orbital with substantial Rydberg character. This is also the case for DEA to formic acid in the 6-9~eV energy range, as confirmed by the scattering calculations described below.

We employ standard electronic structure methods to compute the energies of the relevant neutral and anion states, using multi-configurational self-consistent-field (MCSCF) and multi-reference configuration-interaction (MRCI) techniques. Some care is needed to obtain a balanced description of a negative ion resonance relative to its parent neutral state which can be sensitive to the choice of molecular orbitals employed. We have found that state-averaged MCSCF orbitals based on the (triplet) excited neutral states which are parents of the resonance anion states form a good basis for characterizing the resonances as well as the excited target states.

\begin{table} [b]
	\small
	\begin{center}
		\caption{Electronic excitation energies of the neutral HCOOH states included in the complex Kohn scattering calculations. Calculations performed at the equilibrium geometry of ground-state HCOOH. Values in parentheses are theoretical MRCI results from ref.~\cite{Lourderaj}}
		\begin{tabular}{cccc}
			\hline
			\hline
			Channel  & Symmetry  & Configuration  &Excitation Energy (eV) \\
			\hline
			\\
			1 &     $^1$A$'$     & (1-10a$'$)$^2$(1-2a$''$)$^2$&  0   \\
			2 &      $^3$A$''$    &    10a$'$3a$''$(n$_0\pi$*)  & 6.60 (5.64) \\
			3  &      $^3$A$'$    &    2a$''$3a$''$($\pi\pi$*)  &  7.03 (6.70)   \\
			4 &      $^1$A$''$    &    10a$'$3a$''$(n$_0\pi$*)   &  7.10 (5.96)  \\
			5 &      $^3$A$'$    &    10a$'$11a$'$(n$_0\sigma$*)   &  7.56 (7.42)\\
			6 &      $^1$A$'$    &    10a$'$11a$'$(n$_0\sigma$*)   &  7.68 (7.56) \\
			7 &      $^3$A$''$    &    2a$''$11a$'$($\pi\sigma$*)   &  9.31  \\
			8 &      $^1$A$''$    &    2a$''$11a$'$($\pi\sigma$*)   &  9.40  \\
			\hline							
			\hline
		\end{tabular}
		\label{tbl:energies}
	\end{center}
\end{table}
The resonance positions and widths are obtained from multi-state close-coupling calculations using the well-established complex Kohn method, which has been described previously~\cite{rlm95}. {Table~\ref{tbl:energies} lists the target energies of the 8 lowest states of neutral HCOOH that were included in the complex Kohn scattering calculations to be described below. These states were obtained from state-averaged MCSCF calculations including the ground-state and the first two triplet A$'$ states, with relative weights of 0.43, 0.14 and 0.43, respectively. The calculations were done by doubly occupying the first ten orbitals (9a$'$, 1a$''$) and including five orbitals (3a$'$, 2a$''$) in a complete active space (CAS) MCSCF.}

The trial wave function for the scattering calculations used here takes the form
\begin{equation}
\label{eq:Psi}
\Psi^-_{\Gamma_ol_om_o}=\sum_{\Gamma}\hat{A}(\chi_{\Gamma}F^-_{\Gamma\Gamma_o})+\sum_id^{\Gamma_o}_i\Theta_i.
\end{equation}
The first sum contains the direct product of {N-electron} neutral target states $\chi_{\Gamma}$ and corresponding continuum {orbitals} $F^-_{\Gamma\Gamma_o}${,}  and the second sum runs over {(N+1)}-electron configuration-state functions (CSFs) $\Theta_i$ constructed from bound molecular orbitals. The operator $\hat{A}$ {antisymmetrizes} the {product of} continuum and target wave functions. We emphasize that all energetically open target states, i.e. all excited states up to and including the parent {of the resonance state}, are included in the set {$\chi_{\Gamma}$}. The functions $\Theta_i$ included in the second sum are of two types. The first type consists of all CSFs that can be constructed, consistent with symmetry, from the molecular orbitals used to expand the target state functions. This group of CSFs is necessary to relax strong orthogonality constraints between target and continuum functions and to describe short-range correlation effects. The second group of functions $\Theta_i$ includes all N+1-electron CSFs consisting of N target molecular orbitals and a virtual molecular orbital. This group of terms is essential in describing target relaxation in the presence of an additional electron. Without such terms, the resonance state can appear above, rather than below, its parent neutral state{, and thereby incorrectly appear to be a core-excited shape resonance instead of a narrow Feshbach resonance.}

Resonance parameters are obtained from the scattering calculations by fitting the eigenphase sums to a Breit-Wigner form. We use the computed body-frame S-matrix elements to connect the theoretical results to laboratory-frame angular distributions by computing the entrance amplitude, as described in Refs~\cite{angular, topical}. The entrance amplitude is a complex-valued matrix element of the electronic Hamiltonian between the resonance wave function and a background scattering wave function, the latter characterized by the electron with momentum vector $\mathbf{k}$, with polar angles $\theta$ and $\phi$, incident on the fixed-in-space molecular target:

\begin {eqnarray}
\begin{aligned}
	V(\theta,\phi;\Xi)&=<\Psi_{res}(\Xi)|H_{el}|\Psi_{bg}(\theta,\phi;\Xi)>\\
	&\equiv <Q\Psi|H_{el}|P\Psi>\, ,
\end{aligned}
\end{eqnarray}
where $\Xi$ labels the internal {nuclear} coordinates of the molecule and the integration implied is over the electronic coordinates. When the relative orientation of the fragments is not observed, as is generally the case, the angular distribution of the DEA product ions is given by
\begin{equation}
\label{axial-recoil}
\frac{d\sigma_{\bf{DEA}}}{d\theta}\propto \int d\phi |V(\theta,\phi;\Xi)|^2\, ,
\end{equation}

A proper, direct evaluation of the $PQ$ matrix element is not straightforward~\cite{DomckePQ}. Alternatively, the entrance amplitude can also be defined in terms of the residue of the fixed-nuclei $S$-matrix at the complex resonance energy. Making use of the form of the $S$-matrix near a narrow resonance, as outlined in Ref.~\cite{angular}, we write $S$ as~\cite{macek, brenig}
\begin{equation}
S=S^{bg}+UBU^\dagger \,,
\end{equation}
where $S^{bg}$ is the slowly varying background part of the $S$-matrix and $B$ is a rank 1 Hermitian matrix. In a partial-wave representation, it can be shown~\cite{angular, ammonia} that
\begin{equation}
\label{eq:Sbg+Sres}
S=S^{bg}+UBU^\dagger \,,
\end{equation}
where $U$ is the unitary transformation that diagonalizes $S^{bg}$ and the matrix elements of $B$ are given by
\begin{equation}
\label{eq:B}
B^{\Lambda\Lambda'}_{lm,l'm'}=i\left(\frac{\gamma^{\Lambda}_{lm}\gamma^{\Lambda'}_{l'm'}}{E-E_r+i\Gamma /2}\right)\, .
\end{equation}

The $\gamma^{\Lambda}_{lm}$ are  {\em complex} partial widths describing decay of the resonance into the $(\Lambda,l,m)$ background channel. Note that the background eigenphases have been incorporated into  $\gamma^{\Lambda}_{lm}$.  The unitarity of $S$ demands that~\cite{taylorbook}
\begin{equation}
\label{eq:eigsum}
\Gamma=\sum_{\Gamma ,l,m}|\gamma^{\Lambda}_{lm}|^2\,.
\end{equation}
We thus obtain, in the partial-wave representation, the following expression for the entrance amplitude: 
\begin{equation}
V(\theta,\phi;\Xi)=\sum_{l,m}i^l\gamma^{\Lambda}_{lm}(\Xi)Y^*_{lm}(\theta,\phi)\,.
\end{equation}
We can thus summarize the procedure for determining the full set of parameters needed to determine the entrance amplitude, at a given nuclear geometry, as follows:
\begin{enumerate}
\item{ Carry out fixed-nuclei electron-molecule scattering calculations to obtain the multi-channel $S-$matrix. For this step, we use the complex-Kohn variational method~\cite{rlm95, rmol95}.}
\item{Fit the eigenphase sum to a Breit-Wigner form to obtain the resonance position $E_R$ and width $\Gamma$.}
\item{Obtain the partial widths $\gamma^{\Lambda}_{lm}$ by fitting the $S-$matrix to Eqs.~\ref{eq:Sbg+Sres} and \ref{eq:B}, with $E_R$ and $\Gamma$  fixed using values from step 2. Eq.~\ref{eq:eigsum} is not imposed in the fitting, but rather used to gauge the overall accuracy of the fit.}
\end{enumerate}

\section{Dissociation Dynamics}
\subsection{{Experimental Ion Yields and Electronic Structure Calculations}}
\begin{figure} [h]
	\includegraphics[width=8.9cm, trim= 5.5cm 2.5cm 7.5cm 0.5cm, clip]{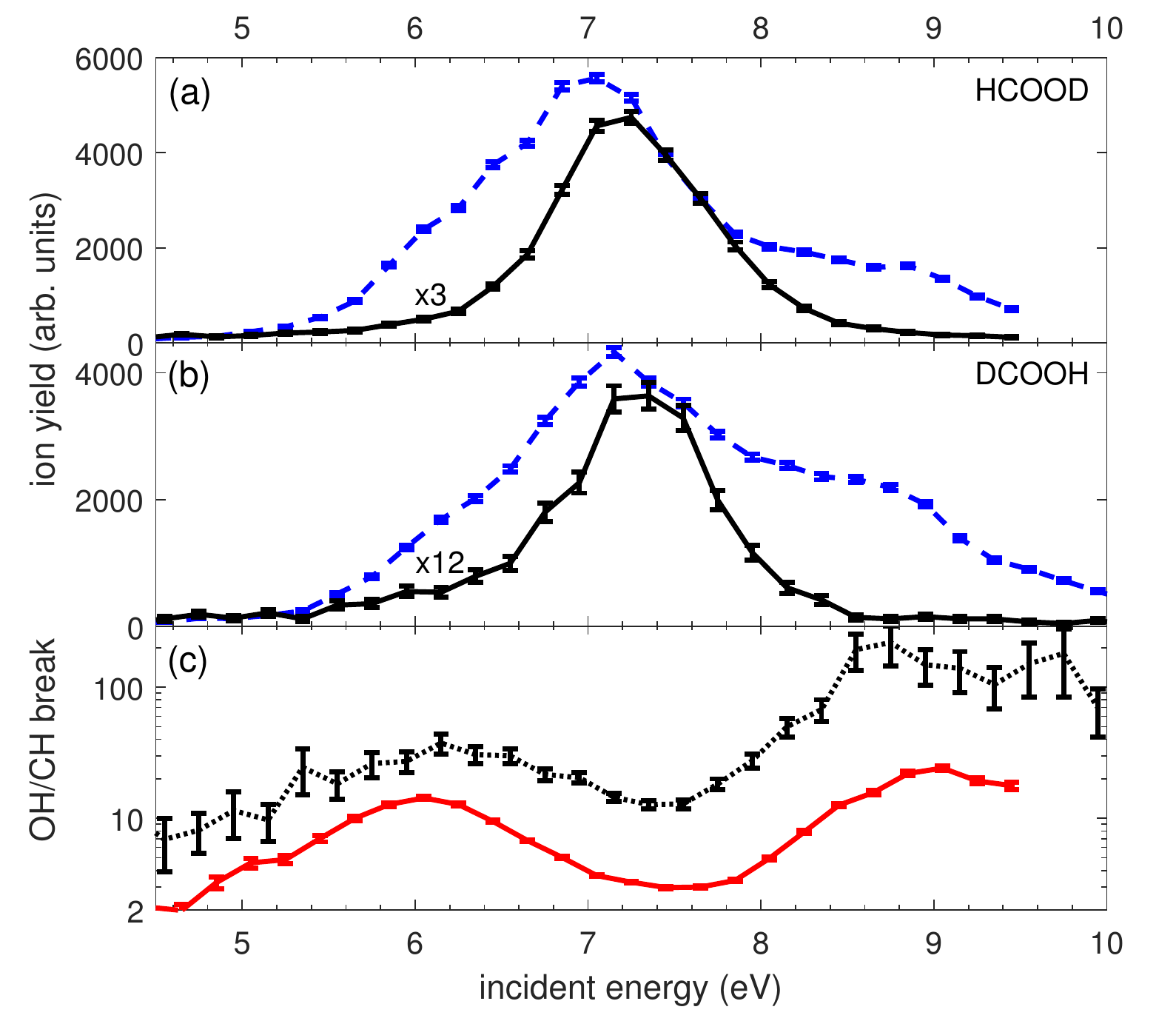}
	\caption{(Color online) Incident electron energy dependence of relative yields of anion fragments for dissociative electron attachment to deutrated formic acid. {(a) C-H break measured by H$^-$ yield (black solid line) and O-D break measured by D$^-$yield (blue dash line) from HCOOD; (b) C-D break measured by D$^-$ yield (black solid line) and O-H break measured by H$^-$ yield (blue dash line) from DCOOH. C-H break and C-D break yields are multiplied by 3 and 12, respectively for clarity. (c) D$^-$/H$^-$ yield from HCOOD (red solid line) and H$^-$/D$^-$ yield from DCOOH (black dotted line), showing the relative DEA yield of O-H break / C-H break for the two formic acid isotopologues.}  
	}
	\label{fig:yields}
\end{figure}

The relative yields of anions produced by C-H and O-H break, measured as a function of incident electron energy are presented in Fig.~\ref{fig:yields}. 
A single  peak in the H$^-$(D$^-$) yield is measured for C-H (C-D) bond scission, having a width of about 1 eV and symmetric shape about its maximum at 7.1 eV (black solid line in Figs~\ref{fig:yields}(a) and (b)). A peak at the same energy also occurs in the O-H break product yield (blue dashed line in Figs~\ref{fig:yields}(a) and (b)), however the shape is considerably different and about 0.5 eV broader than the corresponding C-H peak. A higher energy shoulder in the H$^-$(D$^-$) yield for O-H (O-D) scission is prominent at about 8.5 eV. 

Determining the absolute DEA cross sections from the present anion yields is not possible in the present experiments. However, the relative yields of anion fragments shows that C-H(D) bond breaks are significantly less probable than O-H(D) breaks; there is also a significant isotope effect for bond scission, particularly for C-H(D) bond breaks. This is seen more clearly in Fig.~\ref{fig:yields}(c), comparing the relative ion yield fraction O-H break / C-H break for each isotopologue. Here we see that O-D break contributes at least a factor of 3 more than C-H break to the total H$^-$ (D$^-$) ion yield for formic acid deuterated at the hydroxyl site (HCOOD, red solid line in Fig.~\ref{fig:yields}(c)), yet the relative contribution from O-H break is more than 12 times higher than C-D break for the formyl-deuterated isotopologue DCOOH (black dotted line in Fig.~\ref{fig:yields}(c)).

\begin{figure}[h]
	\includegraphics[width=8.5cm]{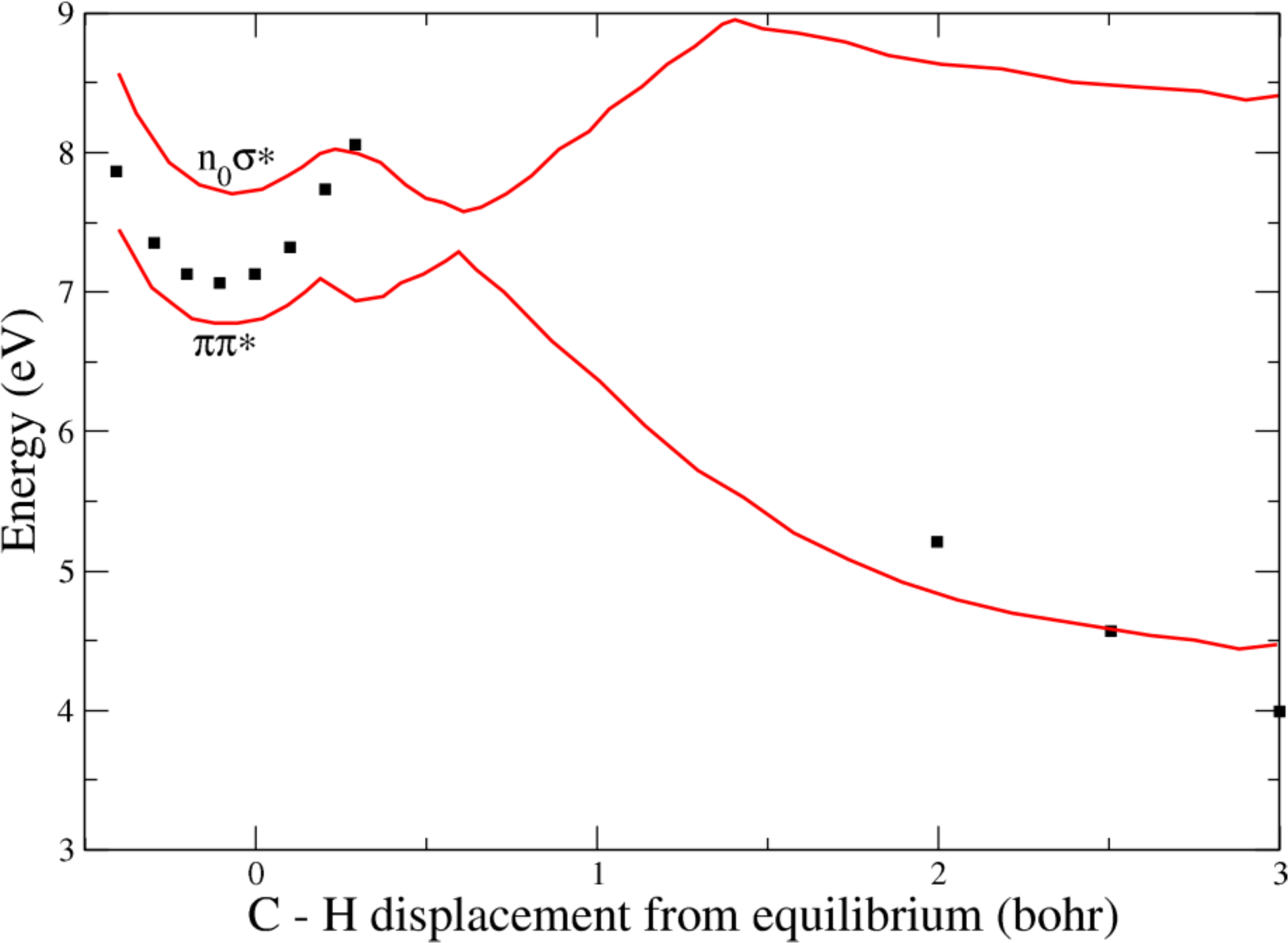}\\\includegraphics[width=8.5cm]{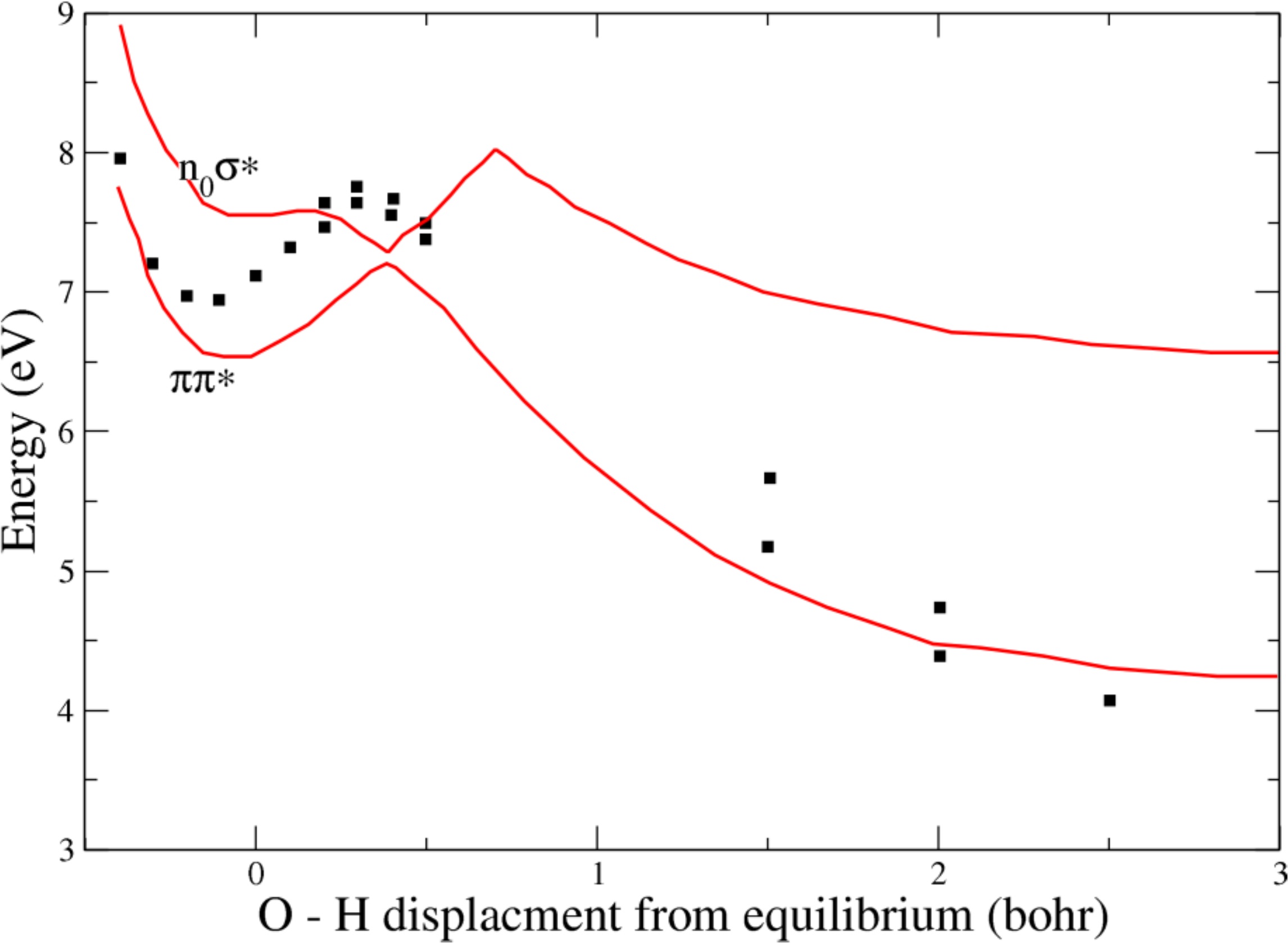}
	\caption{(Color online) Lowest $^3$A$'$ electronically excited states of neutral formic acid (solid curves) and n$_0(\sigma*)^2${, $^2$A$'$} Feshbach resonance (squares) as a function of C-H and O-H displacement from equilibrium. The parent triplets have {conical intersections}, but the resonance can be followed to H$^-$ + HOCO for CH and H$^-$ + HCOO for OH dissociation.} 
	\label{fig:curve-1}
\end{figure}

We turn to the results of electronic structure calculations for an interpretation of the 7.1 eV peak, which figures prominently in both C-H and O-H bond scission. Fig.~\ref{fig:curve-1} shows the electronic energy of the two lowest triplet A$'$ excited states of formic acid as a function of C-H and O-H displacement from equilibrium. 
These curves were obtained from state-averaged MCSCF calculations using the two lowest {$^3$A$'$} states to generate the orbitals. The calculations were done by doubly occupying the first eight orbitals (7a$'$, 1a$''$) and including six orbitals (4a$'$, 2a$''$) {in a CAS-MCSCF.} The resonance energies are also plotted at several geometries. These were obtained by carrying out CAS plus single excitation MRCI calculations on the $^2$A$'$ anions and searching for the eigenvalue whose largest configuration interaction (CI) coefficient corresponded to the doubly-excited resonance state. Near equilibrium geometry, it is the n$_0\sigma^*,^3$A$'$ state which is the parent of the lowest $^2$A$'$ resonance {at 7.1 eV}; this {assignment} was also confirmed by the scattering calculations described below. Evidently, this resonance can dissociate to both H$^-$ + HOCO or H$^-$ + HCOO fragments. We see that the parent triplet states of the Feshbach resonance have {conical intersections}, in particular a crossing with the triplet $\pi\pi^*$,$^3$A$'$ state near 0.5 bohr C-H or O-H displacement, but the resonance can nevertheless be followed to dissociated products. 

Note that there are barriers to dissociation in both cases and that the resonance appears to rise above its parent near 0.25 bohr CH or OH displacement. The latter behavior is undoubtedly unphysical and occurs, on the one hand, because the geometries of the fragments in these calculations were held fixed and not re-optimized at each CH or OH displacement. It is also a reflection of the fact that a better balance between N- and (N+1)-electron correlation is required away from equilibrium geometry. Note that when the resonance appears above its parent neutral state, it broadens and more than one root of the CI on the anion may have the character of the resonance as is seen in Fig. \ref{fig:curve-1}.  {Nonetheless we can see that the} barrier in the case of C-H break is slightly higher than the O-H barrier, which is consistent with the smaller cross section for C-H break relative to O-H break, as well as the observed isotope effect (Fig. \ref{fig:curve-1}(c)). It is noteworthy that in the case of C-H  scission, we find only one channel that correlates with H$^-$ +  {HOCO  -- a} simple reflection of the fact that there are evidently no low-lying excited states of the HOCO radical. On the other hand, as we discuss further below, O-H scission produces the formyloxyl radical, which has three low-lying excited states~\cite{mclean, rauk}. 

\begin{figure} [h]
	\includegraphics[width=9.cm]{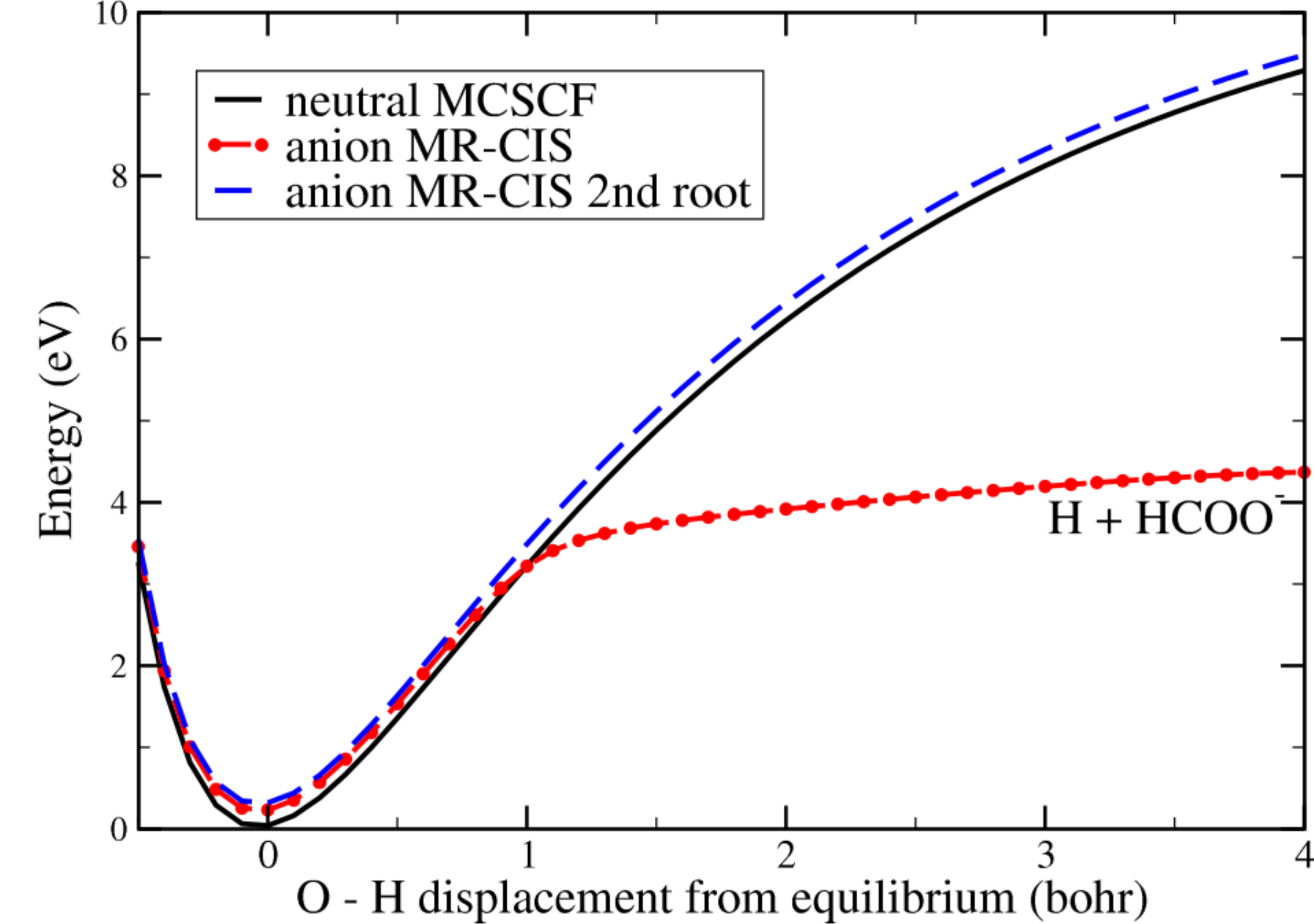}
	\caption{(Color online) Anion potential curve dissociating to the formate anion plus H from multireference singles CI (chained) showing virtual state behavior and second root (dashed) lying in the continuum plotted together with ground state potential curve (solid) from MCSCF whose CAS configurations provide the reference for the CI.  } 
	\label{fig:FormateAnion}
\end{figure}

{Another question about the possible asymptotes of DEA concerns dissociation to produce the formate anion or the HOCO$^-$ anion.   In our electronic structure calculations we found that the lowest anion asymptote correlates with {H + HCOO$^-$}, producing the formate anion.  This channel does not appear to lead to a resonance at small {OH} distances, but rather decreases in energy and becomes a virtual state as the anion energy approaches that of neutral formic acid.  To see that behavior, an electronic structure calculation can be performed that has a degree of consistency between the anion and the neutral as follows.  First an MCSCF calculation is performed on the neutral with 8 a$'$ orbitals doubly occupied and a CAS space of 4 a$'$ and 3 a$''$ orbitals.  Then an MRCI calculation is performed on the anion using the MCSCF orbitals in the reference space with the 8 a$'$ orbitals frozen and only single excitations out of the same CAS space.  Such a calculation using a large Gaussian basis including diffuse functions on all centers (aug-cc-pVTZ \cite{ccbasis_Dunning_1992}) for which the formate anion asymptote is easily identified is shown in Fig. \ref{fig:FormateAnion}.  The lowest  HCOOH$^-$ anion is a $^2$A$'$ state and thus has a strong s-wave component. The typical behavior of a virtual state is seen in which there is no avoided crossing between the virtual state and and any other continuum state as it ceases to be a bound state, collapsing to become the ground state of formic acid with an additional electron in a low energy continuum orbital.  We have verified that the same behavior appears for the lowest anion state when dissociating the C-H bond in such a calculation to produce the HOCO$^-$ anion, and that these two asymptotes are connected adiabatically.  Interestingly, this virtual state behavior has been seen in two other molecules that undergo DEA, carbon dioxide \cite{Sommerfeld_CO2_2004,Moraco2} and ammonia \cite{ammonia}, and there is reason to believe such behavior may be common among polyatomic anions.  However, although these asymptotes correlate adiabatically with virtual states, DEA at low incident electron energies does produce the formate anion, but apparently nonlocal dynamics beyond the Born-Oppenheimer approximation are necessary to understand the mechanism \cite{Fabrikant,Allan_2013}.}

\begin{figure} [h]
	\includegraphics[width=8.5cm]{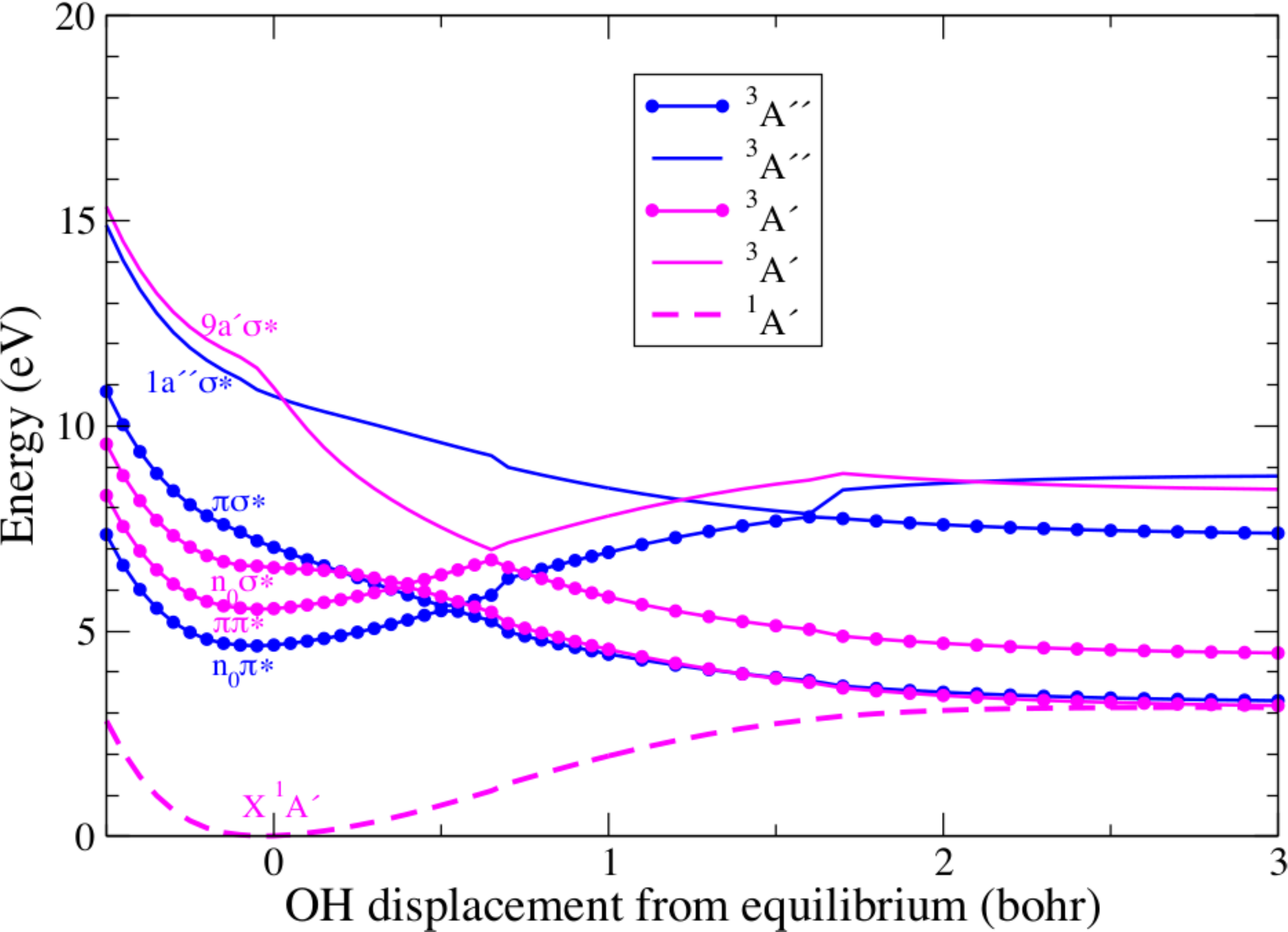}
	\caption{(Color online) HCOOH ground and triplet excited states. Curves with solid circles obtained from 4-state averaged MCSCF as described in text; other states obtained from CAS-CI using MCSCF orbitals.} 
	\label{fig:triplets}
\end{figure}

While C-H scission can only lead to H$^-$ production through the lowest A$'$ Feshbach resonance, there are other excited triplet states of HCOOH that can serve as parent states of Feshbach resonances dissociating to H$^-$ + HCOO*. These are expected to involve single excitation of the 2a$''$, 9a$'$ or 1a$''$ (ie, HOMO-1, HOMO-2 or HOMO-3) orbitals into the $\sigma$* orbital. Figure~\ref{fig:triplets} shows the electronic energy of the ground state of HCOOH and the six lowest triplet states as a function of O-H displacement from equilibrium. These results were obtained by averaging  the two lowest $^3$A$'$ and two lowest $^3$A$''$ states in a CAS-MCSCF calculation and using these orbitals in a CAS-CI for the other states. While a cursory inspection of these curves would suggest that only the n$_0\sigma$*  (2$^3$A$'$) and the $\pi\sigma$* (2$^3$A$''$) states can serve as parents of doubly excited resonance anions in th 6-9 eV region, it should be borne in mind that the HCOO moiety in these calculations is held fixed at the equilibrium HCOOH geometry. At their individual equilibrium geometries the formyloxyl  radicals (HCOO$\cdot$) all have C$_{2v}$ symmetry and the B$_2$ (n$_0^{-1}$) and A$_1$ (9a$'^{-1}$) radicals are only split by $\sim$0.1 eV~\cite{mclean, rauk}. We will return to this point below.

\begin{figure} [h]
	\includegraphics[width=8.5cm]{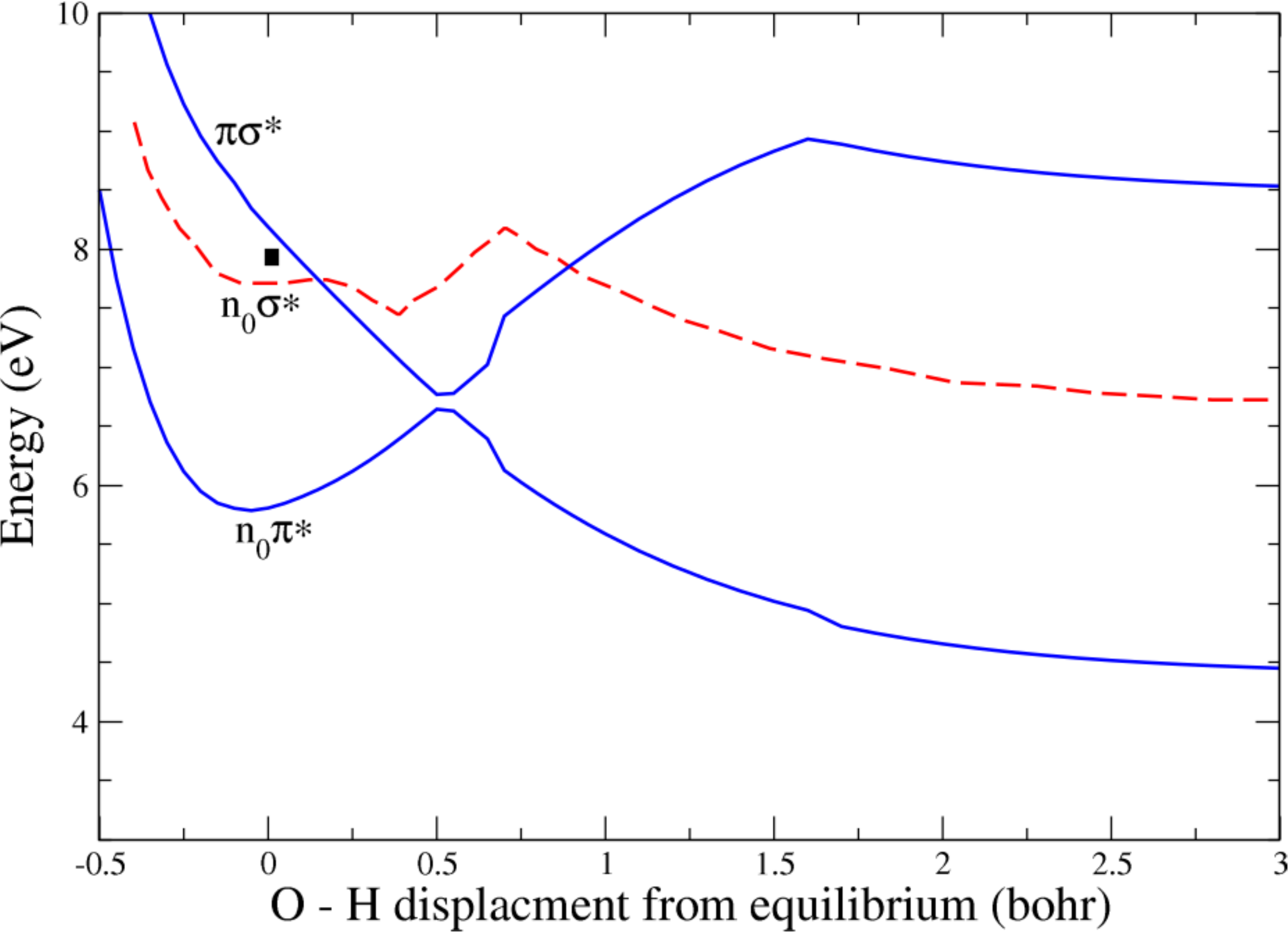}
	\caption{(Color online)  Lowest $^3$A$''$ electronically excited states of neutral formic acid (solid curves) ; the 2a$''(\sigma^*)^2,^2$A$''$ Feshbach resonance (square) is also plotted at equilibrium geometry. The $^3$A$'$ excited state (dashed line) is from {Fig.~\ref{fig:curve-1}b.}} 
	\label{fig:curve-2}
\end{figure}

The first excited triplet state of formic acid,  n$_0\pi^*, ^3$A$''$, does not lead to electron capture into a doubly-excited state. The second excited triplet A$''$ state is produced by a 2a$''$ $\rightarrow$ $\sigma^*$ excitation, which is the parent of a $^2$A$''$ doubly-excited resonance state that dissociates to H$^-$ plus an excited {($^2$A$_2$) formyloxl radical~\cite{mclean, rauk}}. Fig.~\ref{fig:curve-2} 
shows the electronic energy of the two lowest triplet A$''$ excited states of formic acid as a function of  O-H displacement from equilibrium. The $^3$A$''$ curves were obtained from state-averaged MCSCF calculations using the two lowest A$'$ triplet states and the two lowest A$''$ triplet states to generate the orbitals. The CAS MCSCF calculations were done by doubly occupying the first seven orbitals (7a$'$, 1a$''$) and including seven orbitals (4a$'$, 3a$''$) in the active space. These calculations place the 2$^3$A$''$ state approximately 0.5 eV above the 2$^3$A$'$ state near equilibrium geometry, which therefore suggests that a 2a$''(\sigma*)^2, ^2$A$''$ resonance state might be responsible for the 8-9 eV shoulder observed in the H$^-$ (D$^-$) ion yield for O-H (O-D) break in (Fig.~\ref{fig:yields}). 

\subsection{{Angular Distributions}}
\begin{figure}  [h]
	\includegraphics[width=8.9cm, trim= 4cm 0.0cm 4.5cm 0.0cm, clip]{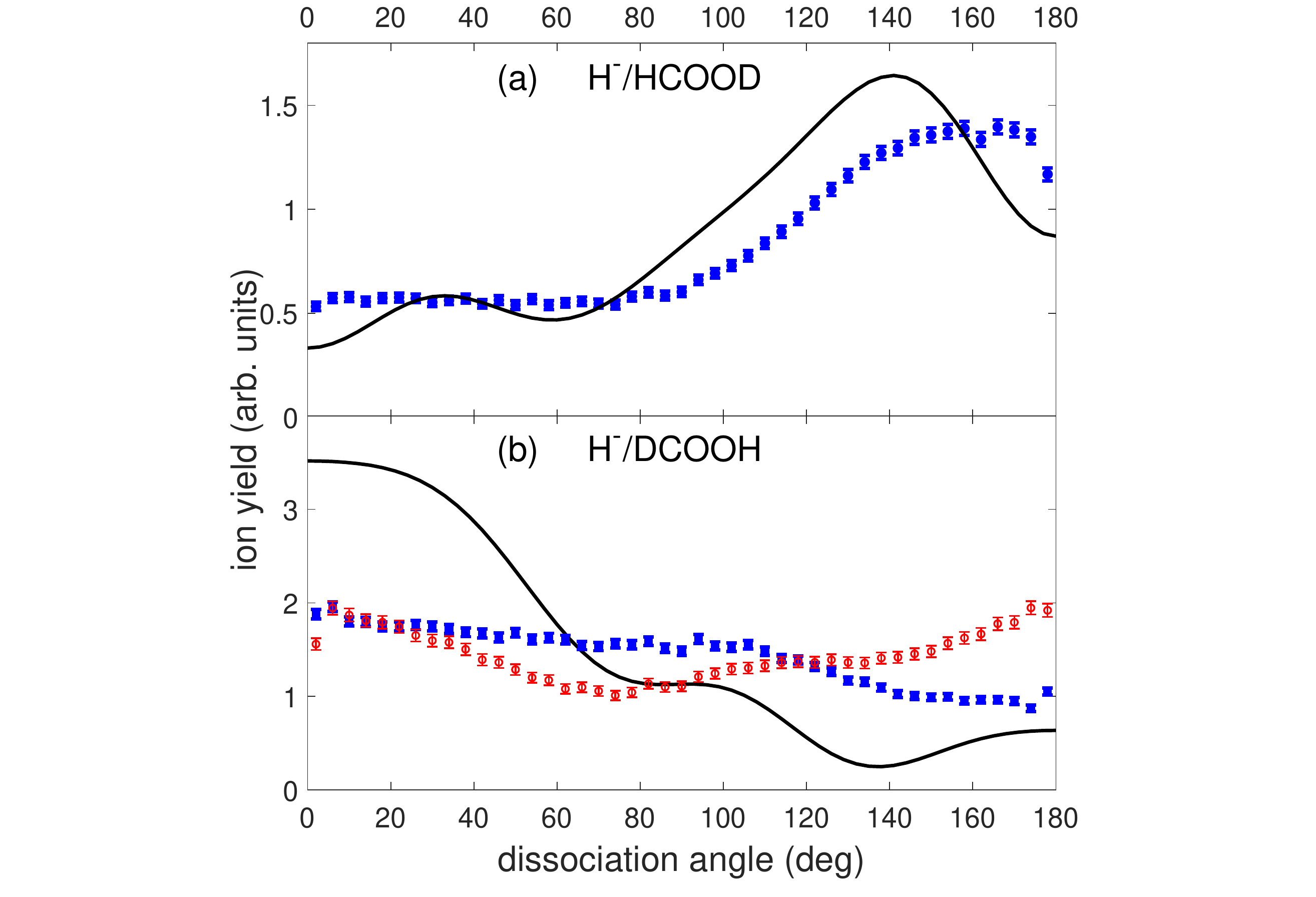}
	\caption{(Color online) Measured and computed H$^-$ ion angular distributions from lowest $^2$A$'$ Feshbach resonance, for (a) HCOOD and (b) DCOOH. The incident energy is 7.25~eV  and the theoretical results computed from entrance amplitude at equilibrium geometry. In panel (b), the blue filled circles are H$^-$ ions with low kinetic energy (1.5 to 3~eV) and the red open circles are H$^-$ ions with high kinetic energy (3 to 5~eV). Error bars show 1 standard deviation in the statistical uncertaintly, and the experimental data are normalized to the theory.}
	\label{fig:a1_angularH}
\end{figure}
\begin{figure} 
	\includegraphics[width=8.9cm, trim= 4cm 0.0cm 4.5cm 0.0cm, clip]{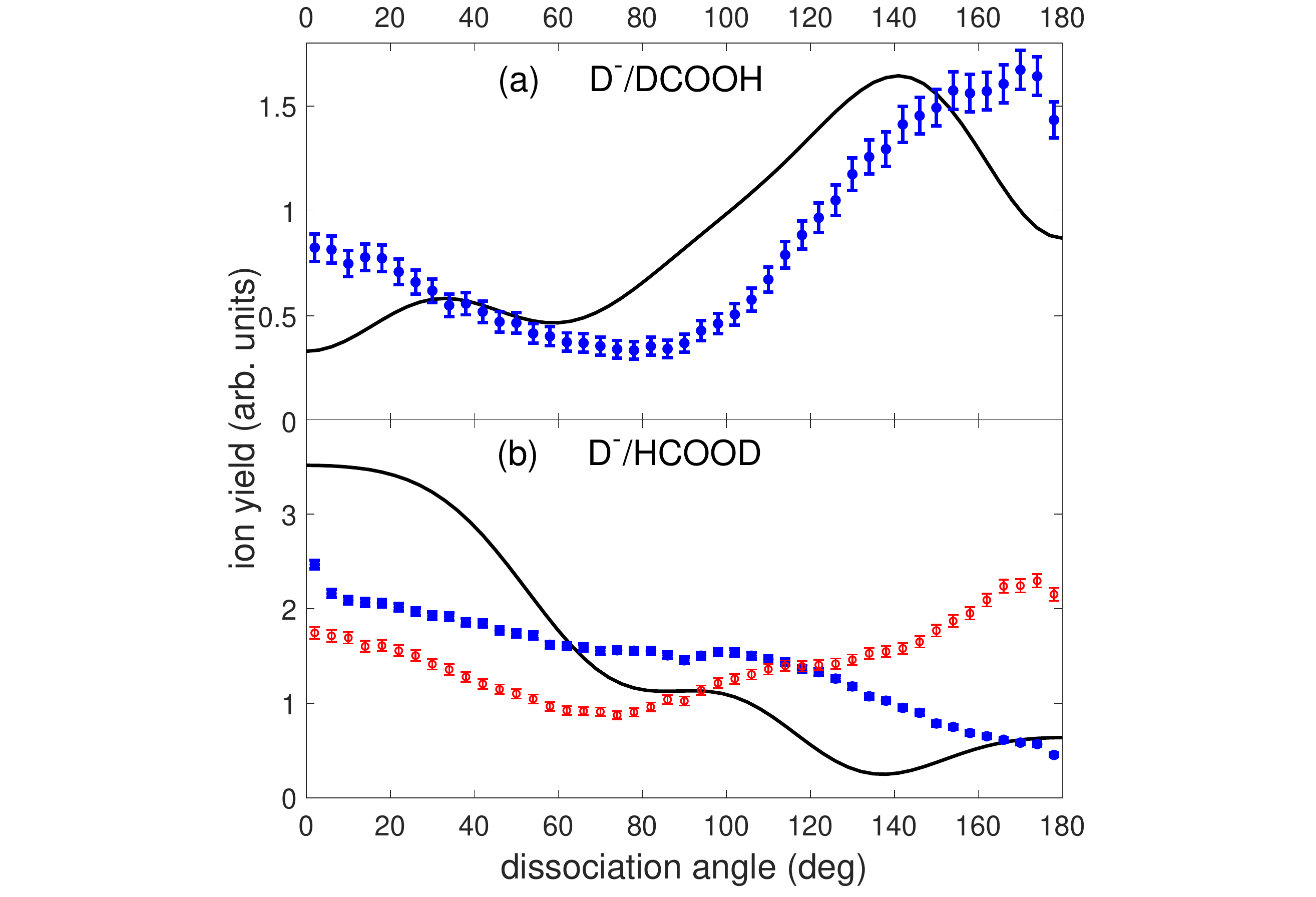}
	\caption{(Color online) Same as Fig.~\ref{fig:a1_angularH}, but for D$^-$ ions from (a) DCOOH and (b) HCOOD. In panel (b), the blue filled circles are D$^-$ ions with low kinetic energy (0.5 to 2~eV) and the red open circles are D$^-$ ions with high kinetic energy (2.7 to 5~eV).}
	\label{fig:a1_angularD}
\end{figure}
\begin{figure}
	\includegraphics[width=9cm]{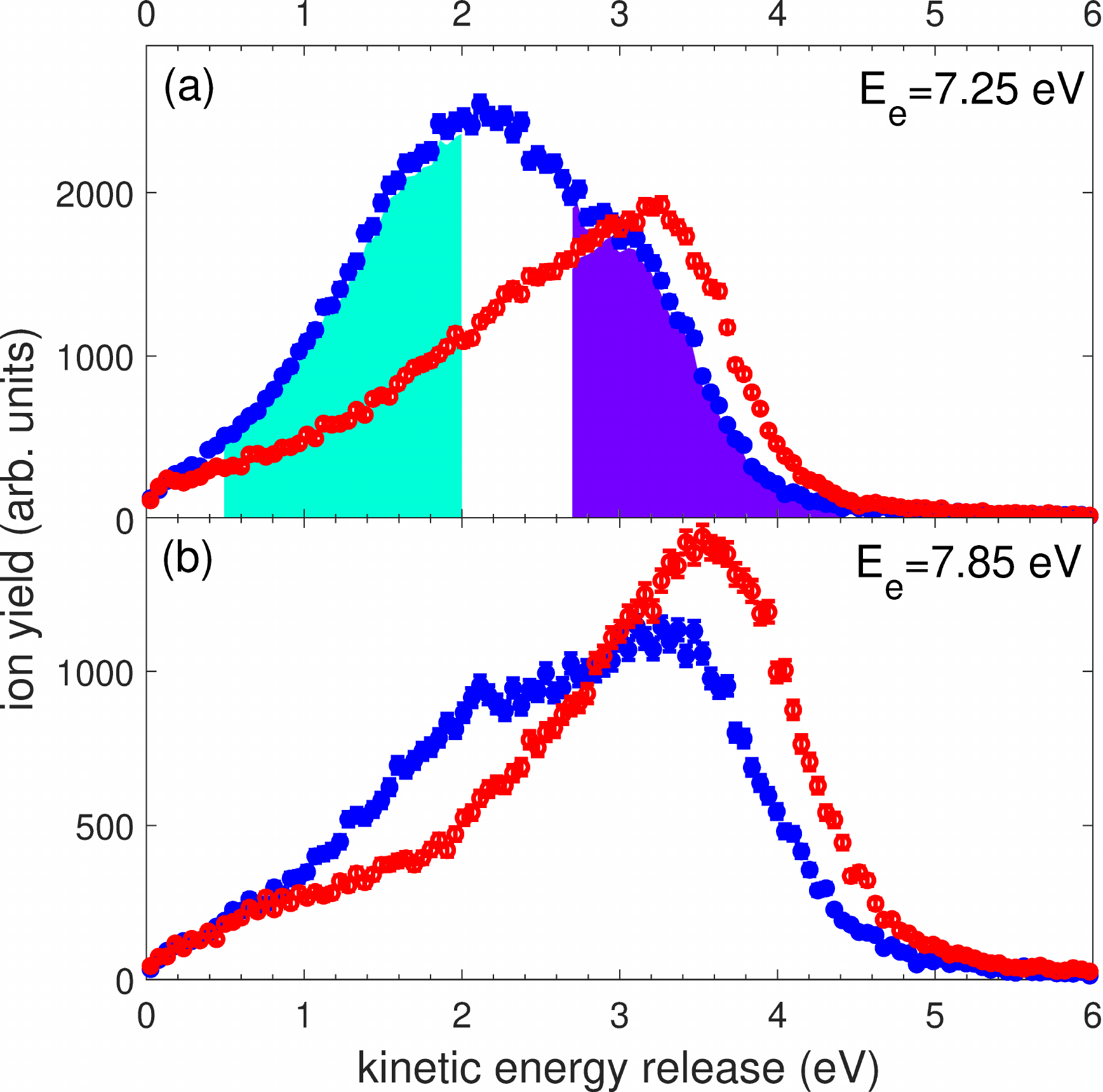}
	\caption{(Color online) Kinetic energy release for D$^-$ + HCOO dissociation from the HCOOD experiment, for ion fragments selected within 45$^\circ$ of the forward (blue) or backward (red) directions. The incident energies are 7.25~eV (top) and 7.85~eV (bottom). The shaded regions (top) indicate the high (magenta) and low (cyan) kinetic energy ranges selected for the angular distributions of Figs~\ref{fig:a1_angularH}(b) and \ref{fig:a1_angularD}(b). Error bars show 1 standard deviation in the statistical uncertaintly.}
	\label{fig:KER}
\end{figure}

Further insight into the breakup dynamics and confirmation of the dissociation channels  assigned above is provided  by a consideration of the measured and calculated H$^-$ angular distributions. The theoretical angular distributions were calculated as described above, using the complex Kohn method to carry out fixed-nuclei scattering calculations at the equilibrium geometry, carrying out a multi-channel S-matrix analysis to extract the partial resonance widths and evaluating the angular distributions from the entrance amplitudes {under the assumption of axial recoil}. We included eight channels in the scattering {calculations}, {using 15 target orbitals (10 frozen, 5 active) calculated as described above in Sec. III}.  The {(N+1)}-electron correlating terms included all configurations that could be constructed from target orbitals in the active space, plus all N+1-terms built from the direct product of N target orbitals and a virtual orbital.
\begin{figure} [h]
	\centering
	\includegraphics[width=0.8\columnwidth,clip=true]{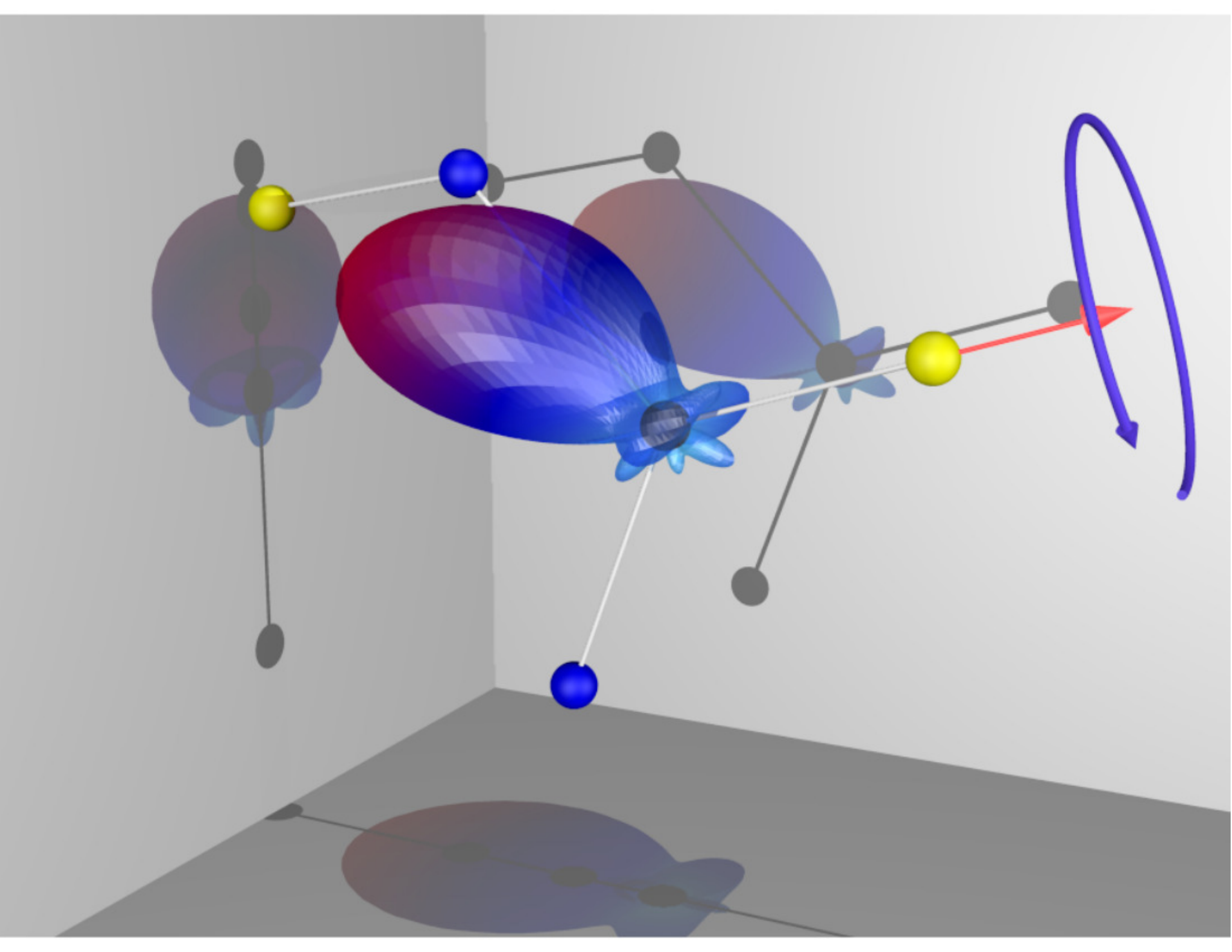}\\\includegraphics[width=0.8\columnwidth,clip=true]{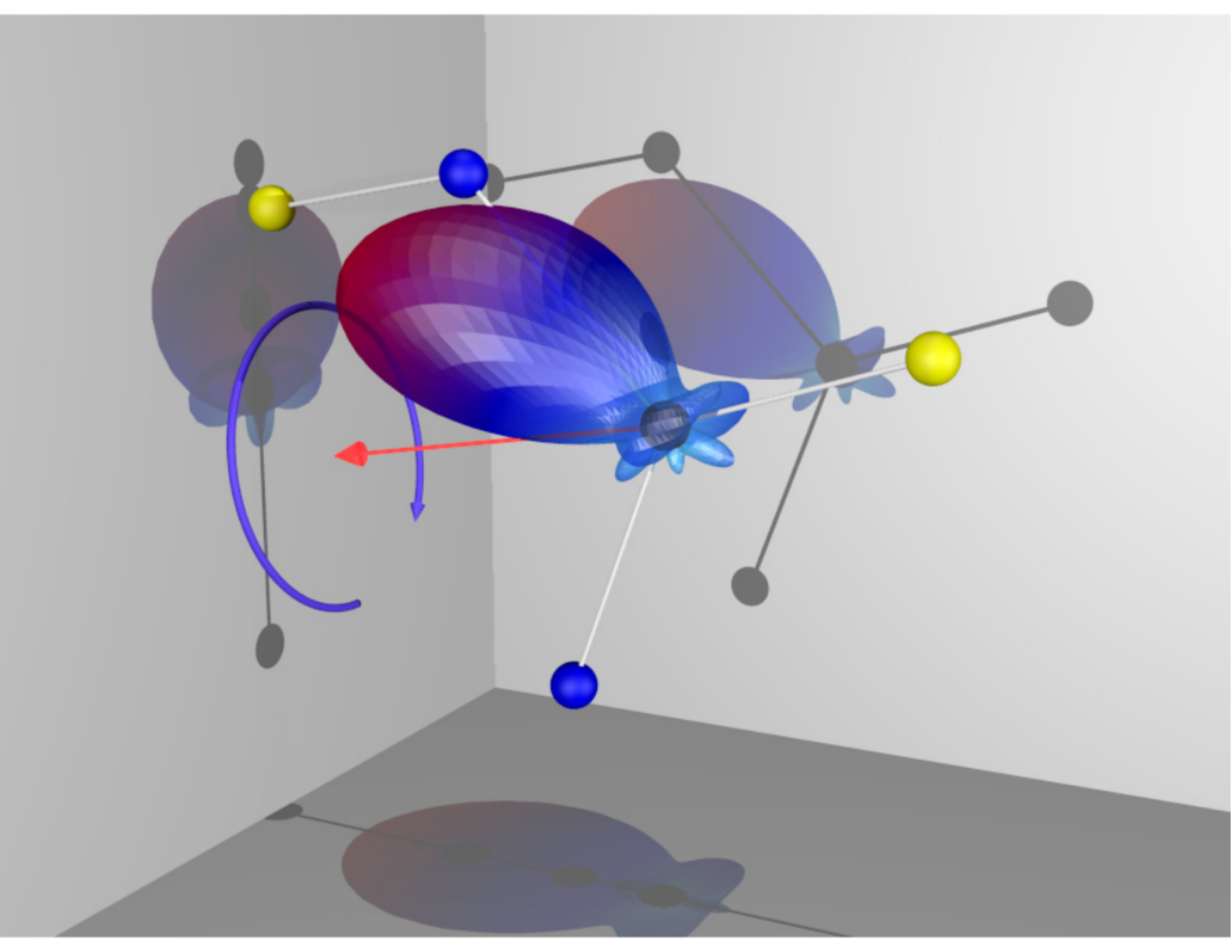}
	\caption{(Color online) 3D entrance probabilities for CH (left) and OH (right) bond scission from lowest $^2$A$'$ Feshbach resonance. Arrows points toward H$^-$ on  C - H and O - H dissociation axes, respectively. } 
	\label{fig:entry}
\end{figure} 

Figures~\ref{fig:a1_angularH} and \ref{fig:a1_angularD} ~{compare the measured} H$^-$ and D$^-$ angular distributions, respectively, at the 7.25 eV resonance for both C-H (C-D) and O-H (O-D) scission with the theoretical result assuming the axial recoil approximation to apply. These channels were distinguished experimentally by using either HCOOD or DCOOH target gases. The angular distributions for O-H (O-D) scission were observed to depend on the ion kinetic energy, so the measured angular distribution for higher kinetic energies (red open circles, 3 to 5~eV for O-H break, 2.7 to 5~eV for O-D break) and lower kinetic energies (blue filled circles, 1.5 to 3~eV for O-H break, 0.5 to 2~eV for O-D break) are shown separately in Figs~\ref{fig:a1_angularH} and \ref{fig:a1_angularD}. These kinetic energy regions are shown in the kinetic energy release (KER) distribution of Fig.~\ref{fig:KER}(a), for D$^-$ ions selected in the forward (blue circles) and backward (red circles) recoil directions, relative to the electron beam direction. For electron incident energies between 7.25~eV (Fig.~\ref{fig:KER}(a)) and 7.85~eV (Fig.~\ref{fig:KER}(b)), the KER distribution consists of two unresolved peaks, most visible in the forward-going D$^-$ fragments (blue circles), that clearly change in relative amplitude as the incident electron energy is scanned over this range. The theoretical angular distributions in Figs~\ref{fig:a1_angularH}, \ref{fig:a1_angularD} were both calculated from the same entrance amplitude, using either the C$\rightarrow$H (C$\rightarrow$D) or O$\rightarrow$H (O$\rightarrow$D) bond vector as the recoil axis in Eq.~\ref{axial-recoil}. From the 3D plots in Fig.~\ref{fig:entry} of the squared modulus of the entrance amplitude (entrance probability), it is clear that the axial recoil approximation predicts backward {(180$^\circ$) and forward (0$^\circ$)} peaked distributions for CH and OH, respectively. It is noteworthy, however,  that there is significantly better agreement between theory and experiment for C-H (C-D) scission than for O-H (O-D) scission. 

Electron attachment energies between 7.25~eV and 7.85~eV were observed to produce structures in the H$^-$ and D$^-$ (Fig.~\ref{fig:KER}) KER distribution, most clearly seen for ion dissociation angles parallel to the electron beam direction (near 0$^\circ$ in Figs~\ref{fig:a1_angularH} and \ref{fig:a1_angularD}). Selecting the high kinetic energy feature, O-H (O-D) dissociation is found to exhibit a structured angular distribution, with broad peaks at 0$^\circ$, $\sim$100$^\circ$ and 175$^\circ$, and a broad minimum around 75$^\circ$. O-H (O-D) dissociation with low KER appears less structured, with the exception of a considerably higher H$^-$ (D$^-$) yield in the forward direction, compared to a much smaller yield in the backward direction. The strong preference for H$^-$ (D$^-$) recoil momentum in the direction approximately parallel to the incident electron, is qualitatively similar to the calculated $^2$A$'$ statem while applying the axial recoil approximation.

\begin{figure} [h]
	\includegraphics[width=8.9cm, trim= 4cm 0.0cm 5cm 0.0cm, clip]{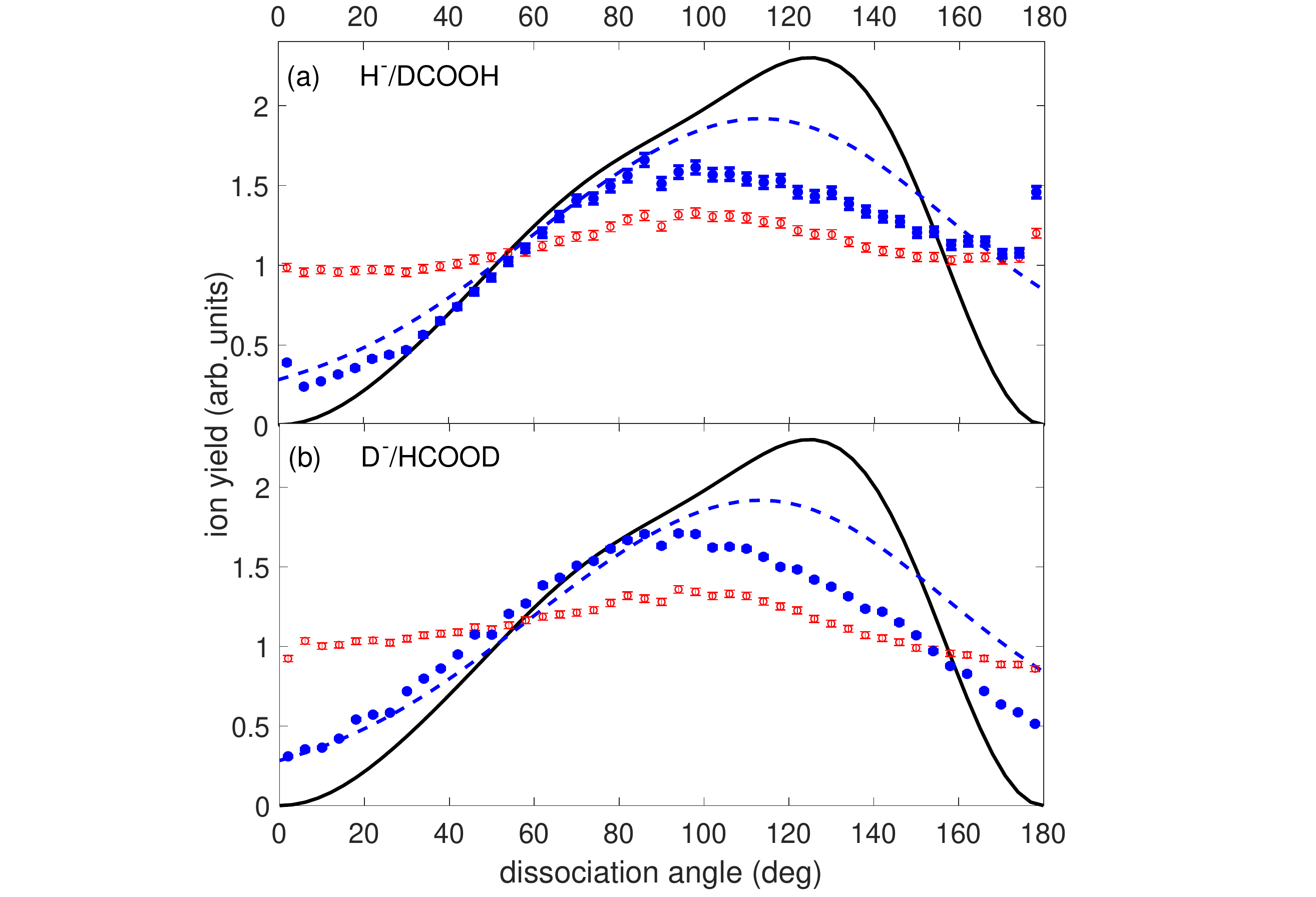}
	\caption{(Color online) Measured (red open circles) and computed (black curve) ion angular distributions from lowest $^2$A$''$ Feshbach resonance: H$^-$ from DCOOH (a) and D$^-$ from HCOOD (b). The incident electron energy is 8.45 eV. Theoretical results computed from 2a$''(\sigma*)^2,^2$A$''$ entrance amplitude at equilibrium geometry, and the experimental data are normalized to the theory. The blue dashed curve shown in both (a) and (b) is a Gaussian convolution (60$^\circ$ full width at half maximum) of the calculated angular distribution to approximate the dissociation dynamics. The blue filled circles are the measured ion angular distribution after subtraction of 50\% of the ion yield from the $^2$A$'$ resonance (measured at 7.25~eV; see Figs~\ref{fig:a1_angularH}(b) and \ref{fig:a1_angularD}(b)). Error bars show 1 standard deviation in the statistical uncertaintly, and the experimental data are normalized to the theory.} 
	\label{fig:a11_angular}
\end{figure}

\begin{figure}
	\includegraphics[width=8.5cm]{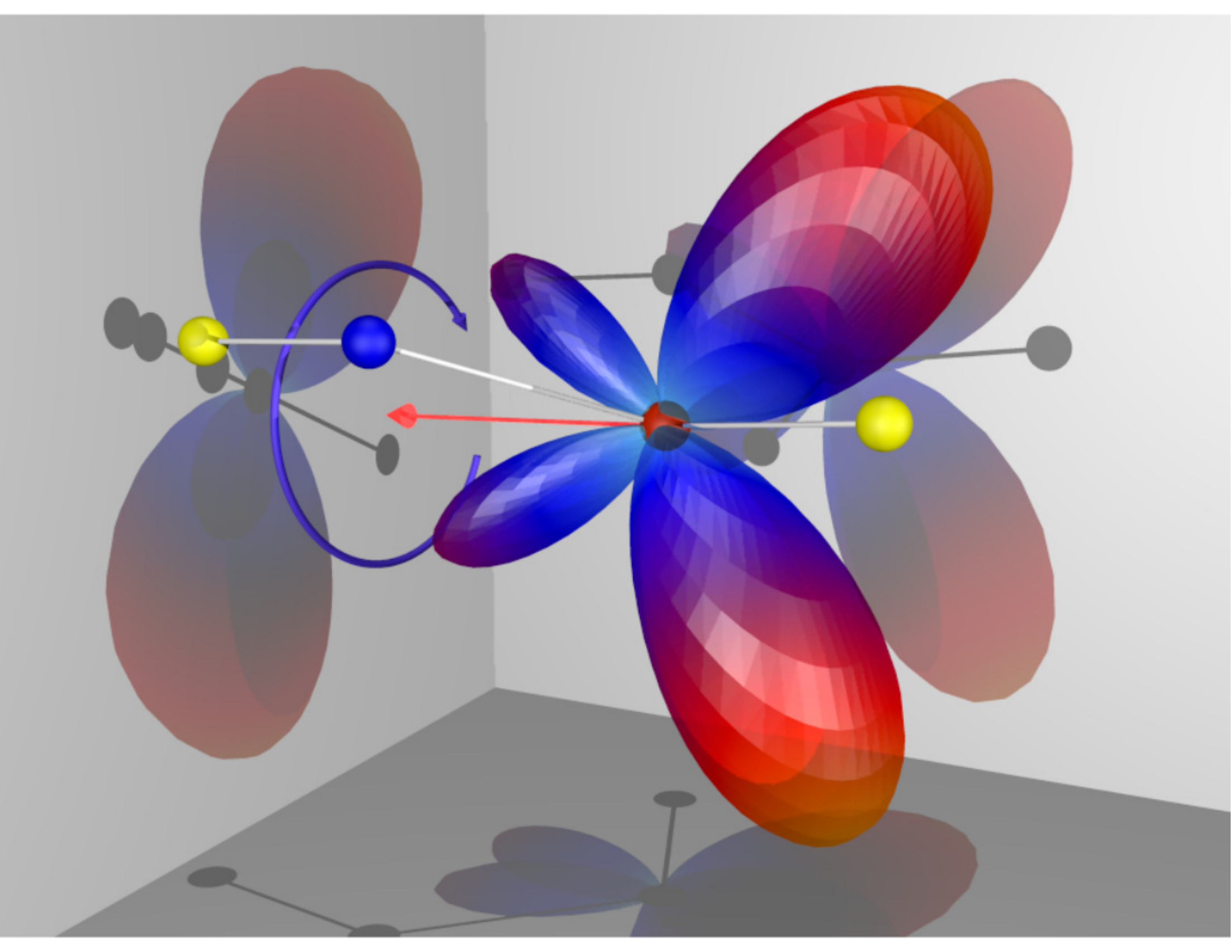}
	\caption{(Color online) 3D entrance probability for OH bond scission from 2a$''(\sigma*)^2,^2$A$''$ Feshbach resonance. Arrow points toward H$^-$ on  O - H dissociation axis.}
	\label{fig:entry-2}
\end{figure}

The structured kinetic energy spectra of Fig.~\ref{fig:KER} suggest two dissociation limits for electron attachment between 7.25~eV and 7.85~eV, {but our electronic structure calculations suggest only the n$_0(\sigma*)^2$, $^2$A$'$ and 2a$''(\sigma*)^2,^2$A$''$ Feshbach resonances participate in DEA between 6 and 9~eV. The observed angular distributions at 7.25 eV are inconsistent with a resonance of A$''$ symmetry, so it is unlikely that involvement of the 2a$''(\sigma*)^2$ resonance can explain the different behaviors measured for low- and high-KER H$^-$(D$^-$) angular distributions resulting from O-H (O-D) bond scission. So if indeed another resonance is implicated in  DEA at 7.25 eV leading to O-H bond scission, it points to the 9a$'$($\sigma$*)$^2$ A$'$ state.  Although our calculations indicate that direct excitation of this resonance below 12~eV electron energy is unlikely (see Fig.~\ref{fig:triplets}),  it is possible that a conical intersection between the 9a$'$($\sigma$*)$^2$ A$'$ state and the lower energy n$_0(\sigma*)^2$, $^2$A$'$ state  could lead to two H$^-$ + DCOO($^*$) (D$^-$ + HCOO($^*$)) dissociation limits, i.e. two electronic states of the formyloxyl radical. Each of these different dissociation reaction pathways could be expected to produce a distinct angular distribution if the potential energy surfaces of the excited formyloxyl radical and the different formic acid anion states have different topologies.
	
	Lastly, we turn to the broad feature centered near 8.5 eV, which is present for OH bond scission but not for C-H bond scission and overlaps the sharper 7.1~eV peak (see Fig.~\ref{fig:yields}). We believe the 2a$''(\sigma^*)^2,^2$A$''$ Feshbach resonance to be responsible for this feature.} Figure~\ref{fig:a11_angular} 
compares the measured H$^-$/DCOOH angular distribution at 8.45~eV with the theoretical distribution for OH scission calculated from the $^2$A$''$ Feshbach resonance. The theoretical results vanish at 0 and 180 degrees, as they must for a resonance of A$''$ symmetry; this can be seen clearly from the entrance probability plotted in Fig.~\ref{fig:entry-2}.
The measurements, on the other hand, show a substantial fragment ion yield at 0 and 180$^\circ$, that could be due to another nearby resonance, having a different symmetry, or rotation of the OH recoil axis during dissociation. It is significant that the broad maximum seen in the angular distribution around 100 degrees is very similar to what is found in the data for OH scission at 7.25 eV (compare Figs~\ref{fig:a1_angularH}b and \ref{fig:a1_angularD}b with Fig.~\ref{fig:a11_angular}). This suggests  that the measurements for H$^-$/DCOOH, at both 7.25eV and 8.45 eV, show contributions from the two overlapping resonances.  We have carried out an approximate method for removing the background A$'$ contribution  from the measured angular distribution at 8.45~eV. Inspection of the ion yields in Fig.~\ref{fig:yields} indicates that roughly 50~\% of H$^-$ from DCOOH at 8.45 eV comes from the lower resonance. Assuming the shape of the A$'$ distribution does not change over the range of energies in question, we then subtract half of the measured distribution at 7.25 eV from the values measured at 8.45 eV. The results are shown as filled circles in Fig.~\ref{fig:a11_angular}.
For comparison, we show the axial recoil calculation and the same calculation with a simple  Gaussian convolution (60 deg FWHM). This procedure significantly improves the agreement between theory and experiment and further supports our interpretation of the dissociation dynamics.

\section {Conclusion}
Production of H$^-$ from DEA to formic acid in the 6-9 eV energy range proceeds via the {direct} excitation of two overlapping resonances. These have been identified as doubly-excited Feshbach resonances involving excitation of either the highest occupied (10a$'$) molecular orbital (HOMO) or the 2a$''$ (HOMO-1) orbital and double occupation of a Rydberg-like $\sigma^*$ orbital. By using deuterated target gases, we have determined experimentally, and verified through {\em ab initio} calculations, that the lower ($^2$A$'$) resonance can produce H$^-$ either through C-H or O-H bond scission, while the upper resonance produces H$^-$ plus excited formyloxyl only through O-H bond scission. 

\begin{figure} [h]
	\includegraphics[width=9cm]{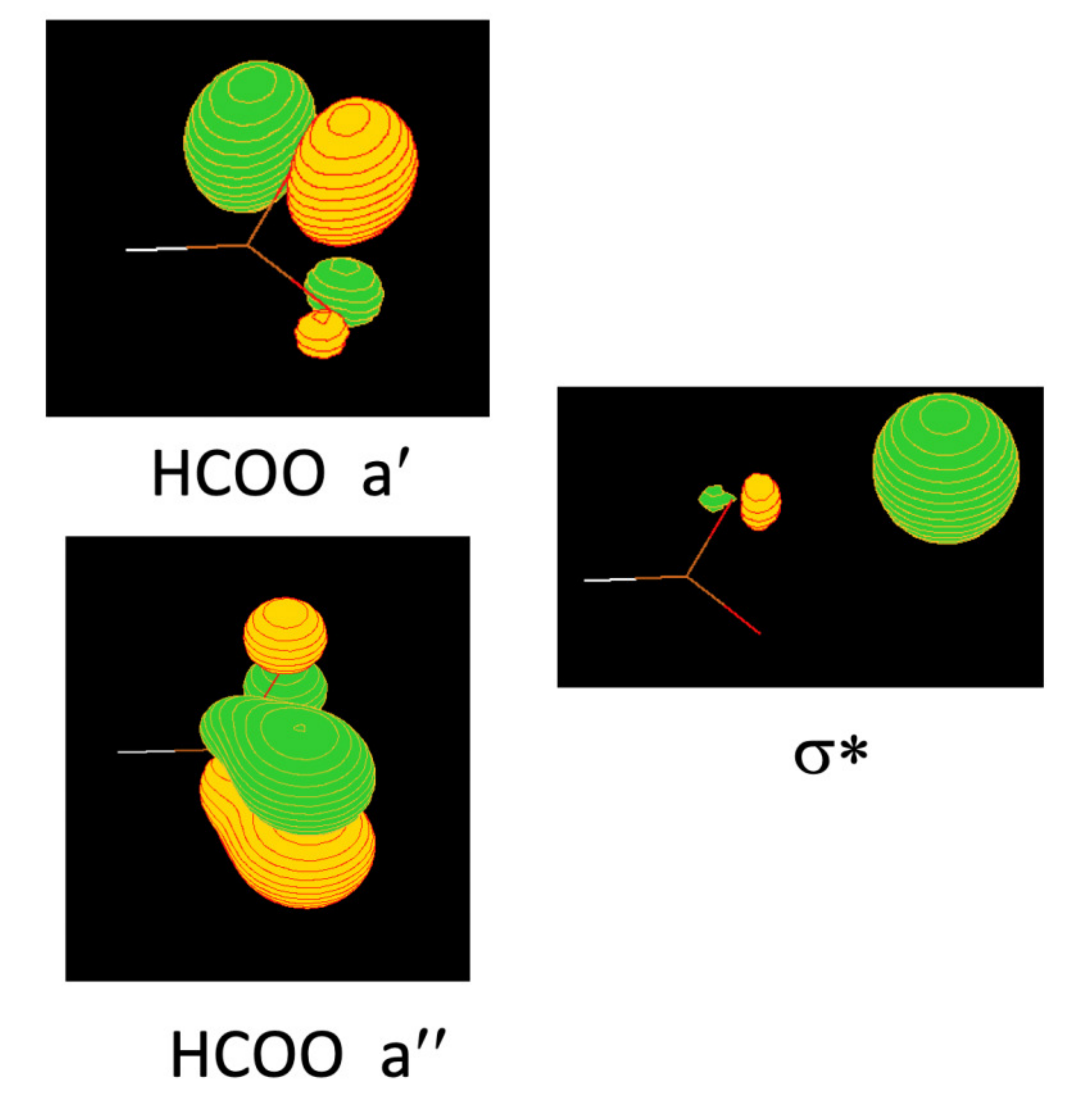}
	\caption{(Color online) Plots of  A$'$,  A$''$  and $\sigma$* molecular orbitals extracted from resonance anions at an O-H separation of 2.5 bohr (see text).}
	\label{fig:molden}
\end{figure}

These resonance assignments and dissociation paths can be seen clearly by an examination of the molecular orbital plots shown in Fig.~\ref{fig:molden}.
These plots were made from the natural orbitals
extracted from the A$'$ and A$''$ resonance anions with the OH separation displaced 2.5~bohr from its value at the geometry of neutral formic acid. The $\sigma^*$ orbital, with an occupation of $\sim$1.7, is clearly that of a dissociating H$^-$ ion, while the two orbitals on the left, with occupations close to 1.0, correspond to formyloxyl radicals of A$'$ (top left) and A$''$ (bottom left) symmetry, respectively. The results of a number of electronic structure studies~\cite{davidson, mclean, rauk} have shown that, at their optimized geometries, the ground state of the formyloxyl radical and its three low-lying excited states all have C$_{2v}$ symmetry. The ground state ($^2$B$_2$) and {the $^2$A$_2$ excited }state of formyloxyl are split by $\sim$0.4 eV~\cite{mclean}, with OCO bond angles of 112$^{\circ}$ and 121$^{\circ}$, respectively. The 10a$'$ and 2a$''$ orbital plots shown in  Fig.~\ref{fig:molden} were computed at the neutral formic acid geometry with an OCO bond angle of 116.8$^{\circ}$. It is clear that these orbitals will correlate with the singly-occupied orbitals of the $^2$B$_2$ and $^2$A$_2$ radicals in C$_{2v}$ geometry.

The experimentally observed resonance features are relatively broad - approximately 1.0 eV and 1.5 eV for the lower and  upper resonances, respectively. We must emphasize that these measured DEA features are not determined by the intrinsic fixed-nuclei electronic widths of the resonance, which are on the order of a few tens of milli-electron volts, but rather by the variation of the dissociative resonance energy surface relative to the neutral target state over the Franck-Condon region.

\begin{figure} [h!]
	\includegraphics[width=8.8cm]{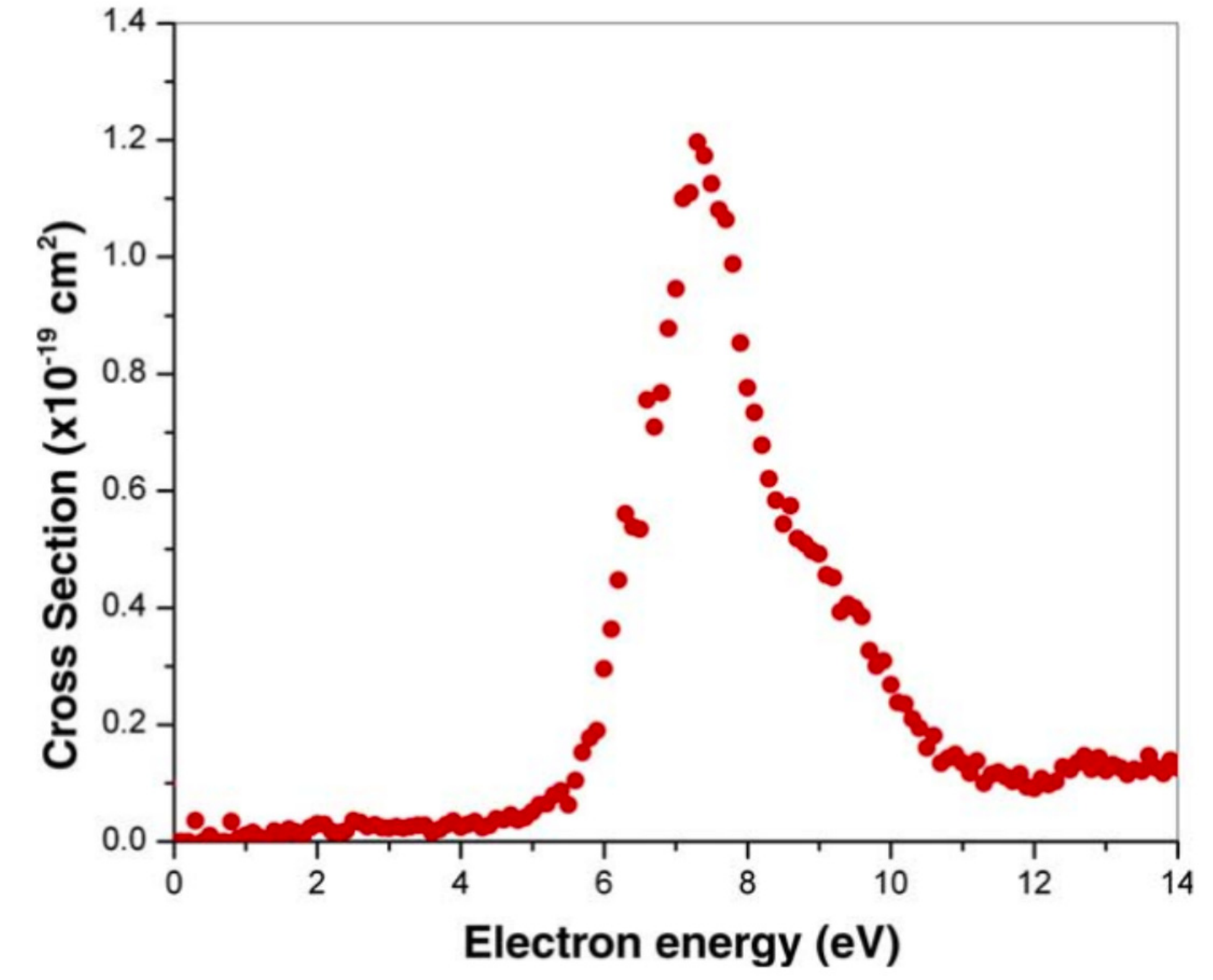}
	\caption{(Color online) Absolute cross section for H$^-$ production from ref.~\cite{Prab05}}
	\label{fig:Prab}
\end{figure}

We conclude with the observation of Prabhudesai {\em et al.}~\cite{Prab05} that the DEA cross section for H$^-$ production, shown in Fig.~\ref{fig:Prab}, 
shows  small features between 12 and 14 eV. Preliminary calculations have indicated that these structures are associated with ($\sigma*$)$^2$ Feshbach resonances that arise from excitation of the inner-valence 1a$''$ or 9a$'$ orbitals of formic acid and correlate with the $^2$B$_1$ or $^2$A$_1$ formyloxyl plus H$^-$ dissociation channels. We have speculated that the 9a$'$($\sigma$*)$^2$ resonance may be indirectly involved in O-H scission at 7.25 eV via a conical intersection with the n$_0$($\sigma$*)$^2$ resonance, but we have yet to study direct excitation of these two higher energy resonances in any detail.

\section*{Conflicts of interest}
There are no conflicts to declare.
\section*{Acknowledgments}
This material is based upon work performed at the University of California Lawrence Berkeley National Laboratory and was supported by the U.S. Department of Energy, Office of Science, Division of Chemical Sciences of the Office of Basic Energy Sciences under Contract DE-AC02-05CH11231.  CST was supported in part by a Berkeley Lab Undergraduate Faculty Fellowship (BLUFF).
\\

\bibliography{hcooh}

\end{document}